*Article*

# Cross-cultural Usability Issues in E/M-Learning

**Mahdi H. Miraz[1,*], Maaruf Ali[2] and Peter S. Excell[3]**

[1]School of Computer Studies, AMA International University BAHRAIN (AMAIUB), Bahrain
m.miraz@amaiu.edu.bh
[2]Dept. of Science & Technology, University of Suffolk, Ipswich, Suffolk, IP4 1QJ, UK
maaruf@ieee.org
[3]Department of Computing, Glyndŵr University, Wrexham, UK
p.excell@glyndwr.ac.uk
*Correspondence: m.miraz@amaiu.edu.bh

**Abstract:** This paper gives an overview of electronic learning (E-Learning) and mobile learning (M-Learning) adoption and diffusion trends, as well as their particular traits, characteristics and issues, especially in terms of cross-cultural and universal usability. E-Learning and M-Learning models using web services and cloud computing, as well as associated security concerns are all addressed. The benefits and enhancements that accrue from using mobile and other internet devices for the purposes of learning in academia are discussed. The differences between traditional classroom-based learning, distance learning, E-Learning and M-Learning models are compared and some conclusions are drawn.

## 1. Introduction

Due to the continual growth of computational power and increasing affordability, the usage of the Internet for learning purposes is becoming a norm. The notion of the "Internet of Everything (IoE)" [1] has inspired people all over the globe to use mobile phones and other hand-held devices for a widening range of internet usage, including that of M-Learning. This paper summarises the literature on deploying E/M-Learning in the field of academia and its adoption trends among students as well as instructors. A sample of more than 50 research papers published in high quality Journals and Conferences were considered and reviewed. Users' adoption trends, usability (especially cross-cultural usability and mobile usability), user interfaces, types of users, technologies and techniques were investigated. Perceiving the next direction of development, deployment, adoption and diffusion of E/M-Learning were the key focus of this review, which was grounded on the following three central themes: 1) Determining the historical background and current state of E/M-Learning; 2) reviewing its usability and adoption trends and 3) systematic forecasting of the future, especially from the present research and trends both in the field of academia and industry. The objective was to devise a specification for a prototype M-Learning package to be developed.

## 2. Electronic Learning (E-Learning)

Although there are many variations of the working definitions of E-Learning (Electronic Learning), it can be simply defined as learning conducted through the use of electronic media and information and communication technologies (ICT), usually on the Internet. In fact, E-Learning in principle covers any sort of electronic educational technology used for both formal and informal learning and teaching purposes. E-Learning consists of or is synonymous with various other terms widely used in academia such as virtual learning environments (VLE), virtual education, technology-enhanced learning (TEL), computer managed instruction, computer aided/assisted instruction (CAI), computer-based instruction (CBI), computer-based training (CBT), internet-based training (IBT), web-based training (WBT), online education, digital educational collaboration and, last but not least, M-Learning. Some of these alternatives only form part of the definition of E-Learning and emphasise on a particular component, aspect or delivery method of it.

E-Learning consists of various types of media, not only text, but also such as images, audio, streaming video or animation. E-Learning may include numerous types of applications, technologies, techniques or processes such as the use of satellite TV, CD-ROM, intranet/Internet, Web 2.0, Web 3.0, podcast and Cloud Computing.

E-Learning can provide greater flexibility by offering pedagogic services anywhere, even outside the traditional classroom. It can be led by an instructor (possibly located at a great distance) or simply be self-paced. Where both tête-à-tête teaching and E-Learning are used together, it is commonly called "blended learning".

## 3. Mobile Learning (M-Learning)

Similarly to E-Learning, although many differing definitions can be found in the relevant literature, Mobile Learning, or M-Learning, may simply be defined as facilitating E-Learning through the use of any mobile or handheld devices. These mobile handheld devices vary in functionality and may include traditional mobile/cell phones, smartphones, PDAs (personal digital assistants), pods, pads (tablets) and other similar devices. Although laptops/notepads are mobile to some extent, they are not considered as "mobile" in this context by many researchers [2]. The principal features and traits of mobile learning have been extensively studied and summarised by Ogata & Yano [3]: "such as situating of instructional activities, permanency, accessibility, immediacy and interactivity". Additional features elucidated by contemporary researchers include: "ubiquity, flexibility, multi-functionality and nonlinearity that mobile devices offer for learning" [4].

M-Learning is seen to be a dictating trend [5] in the domain of emergent technology based applications for academia and is expected to act as a bridge [6], by reducing the gap between traditional and informal learning approaches.

## 4. Seminal Research and Recent Advances of E/M-Learning

The increasing use of mobile technology is influencing cultural practice and facilitates novel contexts for learning [7], although the integration of mobile technologies with traditional teaching has progressed at a somewhat slower rate. In part this is due to the fact that instructors themselves need to be trained in the use of mobile technology [8]. However, as mobile devices and networking technologies are becoming an integral part of daily life, it can be foreseen that M-Learning will soon become more widely adopted by educational programs around the globe.

A study [9] was conducted in 2013 by Olga Viberg and Åke Grönlund to explore the adoption trends among students from different backgrounds, classified by factors such as gender, culture and age. The sample size of the study was 345 students. However, the participants were selected from just two universities, one in China and one in Sweden. The researchers took advantage of "Kearney's pedagogical framework" [10] towards use of M-Learning (mobile learning) from a socio-cultural point of view. Their study [9] also adopted Hofstede's cultural dimensions [11-12] to explore students' cultural views towards mobile learning. Individualisation was found to be mostly positive (83%), followed by collaboration (74%) and authenticity (73%).

The Higher Education sector has to adapt to evolving technologies and their use such as "social media, social networking services, and mobile devices" [13]. Naturally this will have a profound effect on how education is delivered and this will need to be considered carefully [14].

As consumers choose carefully their expensive mobile devices, this often becomes an extension of their personality [15]. Users of these devices are made to create virtual online identities and associated content and interact through social networking media. This phenomenon shows the importance of learning how best to effectively deliver content to such mobile devices to augment these new learning paradigms for end-users [16].

Mobile learning has already permeated and become widespread in higher education. A study [5] by Wu et al. has shown that "85 out of 164 (52%) [as of 2012] of studies about mobile learning have been implemented within a higher educational framework".

Even though the mobile learning industry is mushrooming, the precise factors behind its adoption are still unclear [17-19]. Understanding the core driving factors will help to fine-tune and optimise the strategies and methods that require to be applied for mobile learning for both the stakeholders and the researchers. This is still very pertinent as "our understanding of technology use from the learners' perspective is still quite limited" [20]. There is still much ongoing research to address this uncertainty [17-18, 21].

One major trigger for the adoption of mobile learning is "language". Users whose mother tongue was not the language of instruction used mobile devices for the flexibility it afforded in being able to take the lessons anytime in any place, especially on board different transportation systems [21]. This is a key benefit of mobile devices, enabling users to make productive use of time that would normally be spent in inactivity. One factor that affected the adoption of mobile technology for personal M-Learning in a negative manner was the cost. Other factors with negative influence included the small display dimension of the device, the input capabilities of the device and the ambient environment being a distraction (e.g. being noisy). There was, however, a high degree of user enthusiasm and expectation, with over 66% expressing a desire to use cellular phones for short to long term language learning.

M-Learning adoption intentions are also significantly influenced by the "near-term/long-term usefulness and personal innovativeness". "The perceived long-term usefulness significantly affects the perceived near-term usefulness" [18]. The expectation of novelty has a direct effect on the expected comprehension of the viability of adopting E/M-Learning. Other factors also include the: "perceived behavioural control, attitude and subjective norm" [17].

As students are the main target for the application of M-Learning, their views should be given priority. With this strategic goal in focus, the key factors for the students adopting this technology have been identified to be the: "compatibility of technology with learning styles and needs, availability of encouragement and support from peers and teachers, and learners' attitudes toward technology" [20]. With the internationalisation of the curriculum of universities in their ever expanding markets beyond their home countries, it is important to note that "cultural values [are] important for [the] adoption of ICTs in general" [22-23]. However, this needs further research, especially in correlation to pedagogic mobile technology [23].

The "Technology Acceptance Model" (TAM) [24] is the most popular theoretical model employed for the prediction of the likely take-up of mobile learning. The model has been further refined in its application to gauge differing learning environments by other researchers [18, 23, 25-26]. To further increase the confidence level, two other theoretical models are also employed, these being the: "Theory of Reasoned Action" [27] and the "Theory of Planned Behaviour" [17].

Due to the many advantages of having a powerful computing platform inside the versatile, cheap and small size of the cellular phone, researchers are continually looking at how to best utilise it for language learning [26]. The disadvantages also must not be overlooked in a battery powered device, such as its small screen size and limited screen resolution [28]. The mobile device is also critically dependent on the network coverage and capacity, subject to interference and service outage. In spite of these negative factors, Thornton and Houser [29] showed that cellular phones can certainly be used effectively for providing language learning curricula for students. This concept was further refined to be known as "Mobile-assisted Language Learning" (MALL) and considered one of the most pertinent mobile applications back in 2009 [30-31]. Another aspect of MALL, language learning based on mobile games has been extensively researched by Fotouhi-Ghazvini [32-35].

Other researchers have explored mobile language learning in depth [25, 29, 32, 36-40]. These explorative studies looked at various aspects of language learning and came to the overall conclusion to support the hypothesis that the use of cellular (mobile) technology does indeed effectively "enhance learners' second language acquisition" [2].

**5. Limitations of E/M-Learning**

Although mobile technologies, and hence M-Learning, have numerous advantages such as: mobility, flexibility, user-friendliness, inexpensiveness and accessibility, there also exists some obvious disadvantages. Some major shortcomings of M-Learning include "limited presentation of graphics" [28], small screen display size, lack of cross-cultural usability, inability to easily provide digital-Braille services for the blind or visually impaired user unless specialised hardware is used, high dependency on cellular mobile networks and reception of their signals - which may restrict transmission capacity and face various types of disturbances such as outages.

The Digital Divide [41-42] and Cross-cultural Issues [42-44] such as socio-economic circumstances, cultural background, traditions, beliefs, financial affordability and social and cultural factors also restrict the diffusion and adoption trends of E/M-Learning.

A detailed analysis [42] concluded that cross-cultural issues can no longer be ignored in the globally connected and networked world. Thus, as a result it has become an absolute necessity to address the cross-cultural issues of information systems (IS), reflecting the user behaviours and influencing the way that fixed and mobile broadband technology is being accepted as well as the way it is changing the life styles of different groups of people. Furthermore, a pilot study in Saudi Arabia [45], concluded that the participants were more likely to feel that the Internet or World Wide Web (WWW) were not supportive for cross-cultural exchange because of their linguistic barriers. Thus, the historical, religious, political and education-level factors associated with the user and how under-developed one country is compared to another all need to be addressed and taken into consideration when designing the user interface. The continued emerging markets of the world are predominantly relationship-oriented, the modus operandi being through intricate networks of personal contacts [46]. This has a great impact on on-line shopping and e-business [45] and this factor also needs to be considered.

The shortcomings of cellular technology should not be overlooked and have been classified by Cheon and colleagues [17] as the "users' technical, psychological [47-48] and pedagogical limitations". Technical constraints include the mobile devices' limited display resolution, insufficient memory and international standardisation issues [48-50].

It should also be noted that the human rate of development is often slower and does not keep pace with the rate of technological advances. Pedagogical limitations such as the short attention span of students must never be overlooked, nor the service interruptions experienced when using mobile devices [47, 51]. As mobile devices become part of the lives of a significant proportion of society, the use of such devices for learning, especially language learning [42, 52-54] and for leisure is virtually becoming indistinguishable. Using mobile devices such as tablets for language learning is still an ongoing research effort [26]. Other research [42, 55] suggests that the usage of the Internet will greatly increase if websites are offered in more different languages in a culturally sensitive way.

Apart from technological considerations, it is very possible that there may be commercial constraints on the development path, particularly of software. It is quite likely that vendors of E/M-Learning products may purposely restrict functionalities of their products so that deluxe versions will offer a superior suite of functionalities and ease of use in order to generate more revenue. This could be enabled once people have become dependent on the basic products which are deemed critical to their business operation. These virtual learning products may also be tailored with specific restricted functionalities for certain markets for defence and security reasons, including also the need to control and manage foreign competition.

Releasing products with more features being opened up in a timely and staged manner may also give the illusion to the consumer of product improvement and refinement. Furthermore, E/M-Learning products may be purposely written with features to test for patience and the persistence of the user, this could be especially useful for training the military, the special forces and also for aptitude training.

## 6. A Journey towards Achieving Universal Usability?

One major drawback of E/M-Learning is that it yet cannot be used for blind users, especially those who require using a Braille input/output device. To address this issue, researchers from industries and academia are working towards a Braille e-book. Figure 1 displays a prototype of such a Braille e-book [56-57] that is being developed by a team of researchers at Yanko Design.

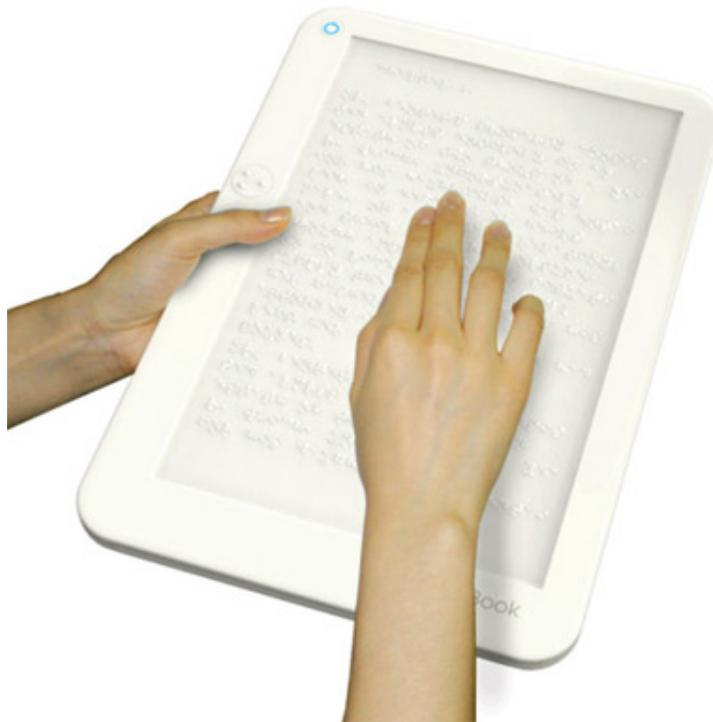

Figure 1:   Braille E-book Concept [56-57].

This device incorporates a "Braille display using electroactive polymers rather than mechanical pins to raise Braille dots on a [grid] display". However, this device is still a concept (2013) and still not deemed to be an economic proposition, unless subsidised for the consumer. A "full Braille computer" was, however, patented as early as 2004, and it continues to be refined and needs further modification for mobile devices.   If these limitations could be eliminated, E/M-Learning is expected to be highly adopted by blind users.

Due to the wide range of diverse users of E/M-Learning, providing cross-cultural usability is a vital demand. A team of researchers [58] from Glyndŵr University, UK and CReATED (Centre for Research on Applied Technology, Entrepreneurship & Development) are jointly working towards achieving this and much progress has been made so far. Identifying the cross-cultural usability issues and integrating some AI (Artificial Intelligence) to alleviate them is the focus of their current research. The project involved designing, developing and testing an M-Learning app to be used by a wide range of users from different backgrounds, initially tested at the University of Ha'il in Saudi Arabia. The test student participants were already familiar with using at least one e/M-Learning app such as Edmodo. Features such as attending exams, reviewing grades and availability of the study materials at any place and time were identified as the most popular ones. Some of them suggested implementing an online version of the Mobile Academy and integrating it with the app. The cross-national (or cross-cultural) aspect of the mLearning app distinguished the work from other examples currently known of.

## 7. Concluding Discussions

E-learning is a concept that has appeared to offer attractive opportunities from the earliest days of computing. Some successes have been achieved, but the pace of adoption has not entirely matched the expectations. The reasons for this are complex, but serve to illuminate related issues, particularly of the human-computer interface. The insertion of the Internet into the system developed into a radical transformation that yielded new avenues for research. Certainly one area where substantial penetration and success have been achieved is in replacement of traditional books by online resources to quite a significant degree. However, interaction with the user, while much useful work has been done, still offers great scope for improvement. This is perhaps illustrated by the observation that the British Open University has been in existence for about 50 years but it has not destroyed the functioning model of traditional universities in the United Kingdom, whose number has grown very substantially over the same period. This appears to illustrate the continuing need for the mentoring function of the human teacher.

A more recent novel factor inserted into the system is the development of mobile learning. Inasmuch as this can be described as a computing device linked to the Internet, it is technologically a very similar concept to E-Learning. However, its ubiquitous presence and the potential opportunity for ownership of a suitable device by the majority of the population of the planet represent a major qualitative change. On the other hand, the severely degraded quality of the mobile human-machine interaction provided (screen and limited input functionality) have a profound effect on the quality of the interaction. Some software products have been developed specifically to accommodate these limitations while still delivering information in a satisfying and agreeable way: this problem is still amenable to the application of much more ingenuity. On the other hand, wearable devices such as Google Glass offer the opportunity for large high-resolution screens to be presented to the user in a mobile device. This product is comparatively new in the market and it remains to be seen whether it will be taken up by a substantial proportion of the population. This will be strongly affected by the rate of production of software that synergistically exploits the capabilities of the device. However, already some public places are either banning or restricting its use.

While mobile learning may well displace many aspects of traditional E-Learning, the need for a human mentor still appears to be as strong as ever and hence M-Learning can only be a supportive tool to the learning process on the human scale. Obviously, it would be economically attractive if this need could be dispensed with, but more research would be needed before this would be likely to occur.

A major problem with any new technology is the inertia of those reluctant to take up new technologies, such as senior academics. Special training needs to be arranged for these people and also possibly special modifications of the user interface to cater for them.

Mobile learning requires a more complex infrastructure such as a reliable local electricity supply and telecommunications network. This implies more points of failure compared to traditional learning modes such as reading a book or problem solving using pen and paper. While less relevant in developed economies, this has implications for the wider adoption of M-Learning in developing nations.

It is also important to look beyond targeting students learners and to consider a focus on senior citizens as the population profile continues to age, especially for the industrialised nations of the globe. Thus it is very important to consider widening accessibility to handicapped, senior and disabled users.

This paper has outlined the domains and characteristics of E/M-Learning, their adoption and diffusion trends, prospects and limitations, as well as current and future research. Possible integration of web and cloud based services with E/M-Learning and associated security concerns were addressed. The benefits and enhancements that E/M-Learning can bring over traditional classroom-based learning were discussed. Universal usability aspects were investigated and Braille devices and applications for facilitating E/M-Learning and cross-cultural usability were explored.

The study undertaken was limited in scope to just over fifty papers which were published in exclusively high impact journals. This does preclude many papers which may have covered many exotic and niche applications, although these niche applications have the potential to one day become common applications.

Thus the lessons learned will be used to design an enhanced prototype application, taking into account the need to match the particular cultural characteristic of the user from the targeted nation with the user interface. The prototype structure needs to appeal to both genders, differing ages, national identities and educational backgrounds of the users and offer an adaptive and adaptable interface to them. The prototype will have core features with additional culturally-appropriate dynamic interfaces that will also adapt with the user. This adaptation will be in terms of the user's learning curve and also the lethargy and/or attentiveness of the user, even suggesting that the user take a break when a session appears to be becoming unproductive, with little progress or even negative progress being made over the lesson duration. This is one particular important lesson that appears to be lacking in the plastic interfaces of the past when applied in an educational context.